\documentstyle[prl,aps,preprint,eqsecnum]{revtex}
\tighten 
 
\begin{document}
\draft

\def\nid{\noindent}
\def\ket#1{|#1\rangle}
\def\bra#1{\langle#1|}
\def\half{\mbox{\small $\frac{1}{2}$}}
\def\quarter{\mbox{\small $\frac{1}{4}$}}
\def\prod#1#2{\langle #1 | #2 \rangle}
\def\mat#1#2#3{\langle #1 | #2 | #3 \rangle}
\def\beq{\begin{equation}}
\def\eeq{\end{equation}}
\def\eeql#1{\label{#1} \end{equation}}
\def\bea{\begin{eqnarray}}
\def\eea{\end{eqnarray}}
\def\eeal#1{\label{#1} \end{eqnarray}}
\def\cotanh{\mathop{\rm cotanh}\nolimits}
\def\cotan{\mathop{\rm cotan}\nolimits}
\def\phs{{\vphantom{*}}}
\def\hn{\mskip-0.5\thinmuskip}
\def\hp{\mskip0.5\thinmuskip}
\renewcommand{\vec}{\bbox}
\def\im{\mathop{\rm Im}\nolimits}

\title{Jordan blocks and generalized bi-orthogonal bases:\\
realizations in open wave systems}

\author{Alec Maassen van den Brink and K. Young}

\address{Department of Physics, The Chinese University of Hong Kong,
Hong Kong, China}

\date{\today}
\maketitle

\begin{abstract}

Dissipative systems can be
described in terms
of non-hermitian hamiltonians $H$, whose
left eigenvectors $\bra{f^j}$
and right eigenvectors $\ket{f_j}$ form a bi-orthogonal system.
Bi-orthogonal systems could suffer from
two difficulties. 
(a) If the eigenvectors do not
span the whole space, then $H$ can only be diagonalized
to blocks (the Jordan-block problem).  
(b) Normalization would not be possible 
and many familiar-looking formulas would fail
if $\prod{f^j}{f_j}=0$ for some $j$ (the orthonormalization
problem). 
Waves in open systems
provide a well-founded realization of a bi-orthogonal
system, and it is shown that these two problems
can indeed occur and are both related to higher-order poles
in the frequency-domain Green's function.  The resolution
is then given
by introducing a generalized duality transformation
involving extra basis vectors, whose time evolution
is modified by polynomials in the time $t$. 
One thus obtains a nontrivial extension of the bi-orthogonal formalism
for dissipative systems.

\end{abstract}

\pacs{PACS numbers: 03.40.Kf, 02.30.Mv, 02.60.Lj, 03.65.-w}
 

\section{Introduction}
\label{sect:intro}

\subsection{Dissipative systems and bi-orthogonal bases}
\label{dissip}

Dissipative systems are often discussed, in a phenomenological way, by
{\em postulating}\/ a non-hermitian hamiltonian (NHH) $H$, whose left
eigenvectors $\bra{f^j}$ and right eigenvectors $\ket{f_j}$ form a
bi-orthogonal system (BS) \cite{wong,rott,faisal,dattoli,sun93}. It is
usually {\em assumed}\/ that these eigenvectors are complete; the BS
then constitutes a bi-orthogonal basis (BB).  These NHHs with discrete
BBs can sometimes be obtained from a full microscopic theory, but
usually under some approximations \cite{faisal,others}.  BBs allow
dissipative systems to be placed into a familiar framework; the
advantages are obvious and need not be enumerated.  As a consequence,
there is a substantial literature on both the mathematics and the
physical applications of BBs (e.g., \cite{mcdouall} (chemical bonding),
\cite{chu}, \cite{hackl} (solid mechanics)).

To introduce the formalism \cite{wong}, consider a Hilbert space with
an NHH $H$, supposed to have a complete basis of eigenvectors
$\ket{f_j}$, with eigenvalues $\omega_j$. In terms of the standard
inner product $\prod{\cdot}{\cdot}$, introduce the unique dual basis
$\{\ket{f^j}\}$ with
\beq
  \prod{f^j}{f_k}=\delta_{jk}\;;
\eeql{dualbasis}
one finds
$\prod{f_k}{H^{\dagger}|f^j}={\prod{f^j}{H|f_k}}^{\!*}=
\omega_k^*\delta_{jk}^\phs$,
so that $H^{\dagger}\ket{f^j}=\omega_j^*\ket{f^j}$. Now define a
duality transformation ${\cal D}$ by ${\cal D}\ket{f_j}=\ket{f^j}$,
extended to the whole space by conjugate linearity~\cite{noDdagger}:
\beq
  {\cal D}  \left( \alpha  \ket{\chi} + \beta \ket{\psi} \right)
  =  \alpha^* {\cal D} \ket{\chi} + \beta^* {\cal D} \ket{\psi}\;.
\eeql{eq:dprop}
The symmetry of (\ref{dualbasis}) shows that the duality transformation
satisfies ${\cal D}^2=\openone$, justifying its name~\cite{c-norm}.
Since $\prod{f_k}{{\cal
D}H|f_j}=\omega_j^*\delta_{jk}^\phs=\prod{f_k}{H^{\dagger}{\cal
D}|f_j}$, one has
\beq 
  {\cal D}H = H^{\dagger}{\cal D}\;.
\eeql{eq:dhhd}

Any vector can now be expanded as
\bea
  \ket{\phi} &=& \sum_j a_j \, \ket{f_j}\label{eq:exp1}\;, \\
  a_j &=& \frac{ \prod{f^j}{\phi} } { \prod{f^j}{f_j} }\;,
\eeal{eq:proj}
where for application in the following Sections it is convenient not to
fix the normalization $\prod{f^j}{f_j}$ by (\ref{dualbasis}). This
leads immediately to the resolution of the identity and of the
time-evolution operator
\bea
  \openone &=& \sum_j \frac{ \ket{f_j}\bra{f^j} } { \prod{f^j}{f_j}}
  \label{eq:comp1}\;, \\
  e^{-iHt} &=& \sum_j \frac{ \ket{f_j} \,  e^{-i\omega_jt} \, \bra{f^j} }
  {\prod{f^j} {f_j}}\;,
\eeal{eq:hrep}
which in principle solves all the dynamics.  For example, the
wavefunction at time $t>0$ with (\ref{eq:exp1}) as initial data would be
\beq
  \ket{\phi(t) } = \sum_j a_j e^{-i\omega_j t} \, \ket{f_j}\;.
\eeql{eq:expt}
It is emphasized that (\ref{eq:dhhd})--(\ref{eq:expt}) have been
obtained only for the case in which $\{\ket{f_j}\}$ constitutes a
complete basis.

This brief account exposes two related lurking difficulties that are
usually brushed aside: the Jordan block problem and the
orthonormalization problem.

\subsection{Jordan blocks and orthonormalization}
\label{jb-intro}

Because $H$ is not hermitian, there is no guarantee
that its eigenvectors $ \ket{f_j} $
form a complete set, so a BB
may not exist.  There are in fact
two fundamentally different reasons why the eigenvectors
may fail to be complete~\cite{c-spec}.  

The first reason can be characterized
heuristically by saying that each eigenvector with a complex
$\omega_j$ is a discrete resonance, and there may be ``background"
under the resonances.  A more formal statement is that
there may be contributions to the Green's function other than
from the discrete poles $\omega_j$ in the complex frequency
plane; this possibility will be further explained below,
but will not be the focus of the present paper.

The second scenario is more intriguing---even though the discrete poles
may give all the dynamics, the associated eigenvectors may nevertheless
be incomplete.  The simplest case is for two (in general any $M \ge 2$)
discrete poles to coalesce into a double (in general $M$th-order) pole,
e.g., upon tuning some system parameter(s).  We shall show below that,
at least for 1-d wave systems, an $M$th-order pole is {\em not}\/
associated with $M$ degenerate eigenvectors, but with only {\em one}.
With the number of eigenvectors reduced by $M-1$, $\{ \ket{f_j} \}$
must become incomplete, and $H$ can only be reduced to block form (in
this case an $M \times M$ block) rather than diagonal form.  This is
known as the Jordan block problem \cite{fnjordan}.  If it were to
occur, the above formulas (\ref{eq:dhhd}), (\ref{eq:hrep}), and
(\ref{eq:expt}), derived assuming an {\it eigenvector\/} basis, would
not hold. Such problems can be handled as the limiting case of several
nearby simple poles, but since some eigenvectors are ``lost", the limit
is singular and needs to be treated with care.

For a non-hermitian operator, in general there is no positivity
condition to guarantee that $\prod{f^j}{f_j}\ne0$ for its left and
right eigenvectors $\bra{f^j}$ and $\ket{f_j}$ corresponding to the
same eigenvalue $\omega_j$, even if both these vectors are unique up to
a constant; in Section~\ref{dissip} this problem did not occur only
because $\ket{f^j}$ was orthogonal to all other eigenvectors
$\ket{f_k}$ ($k\neq j$) of a set which was supposed {\em complete}. If
$\prod{f^j}{f_j}$ should vanish for some $j$, then the above formulas
would also fail.  This is the orthonormalization problem.  (So long as
$\prod{f^j}{f_j}\ne 0$, one can adopt a normalization convention that
it is unity.)

\subsection{Outline of paper}

In most applications, it is implicitly {\em assumed}\/
that the Jordan block problem does not occur,
and the BB formalism has not been extended
to handle such problems if and when they do.
In part, this is because phenomenological NHHs 
do not exhibit these problems
in a natural and convincing way, through which a possible
extension of the formalism could be explored.

Recently it has been shown that waves in certain 1-d open systems
provide an {\em exact}\/ realization of BBs \cite{bior}.  In these
systems, dissipation occurs by the leakage of waves 
out of the system (through the imposition of the outgoing
condition at the boundary).  Under some simple conditions,
the dynamics is completely controlled by the discrete poles
in the frequency plane \cite{comp1,comp2} and an analogy with
conservative systems
can be developed \cite{tong}, such that the formalism
can be cast in the language of BBs.  In terms
of these physical realizations, we shall show that the question
at hand can be reduced to investigating the existence
and properties of higher-order
poles in the frequency-domain Green's function.

In Section \ref{sect:waves},
waves in open systems are described, focusing on the
representation of the dynamics by the discrete poles
\cite{fncomp} and the relation to a BS.  We shall deal with 
both the wave equation and
the Klein--Gordon equation.  The duality
transformation, central to the BS formalism, is shown
to arise naturally from the dynamics.  In Section \ref{sect:hp},
we construct examples where higher-order poles do occur;
the examples correspond to the critical
damping of an oscillator.
We then 
show that these higher-order poles relate 
to the Jordan block problem (``losing" one or more eigenvectors
from a finite subspace).
But since the dynamics is 
contained in the Green's function, a systematic study of the
residue at a higher-order pole should reveal how the 
situation must be handled. We therefore first analyze, in Section
\ref{sect:jb}, the finite-dimensional subspace associated
with the higher-order pole, i.e., the matrix representation of one
Jordan block.
Subsequently, in Section \ref{sect:dual}
we solve for the field evolution, which leads to an extension of the BB
formalism
with a generalized duality transformation.
Then, in Section \ref{sect:degpt}, the
perturbation theory associated with such a block
is presented\cite{kato}.  Unlike
the degenerate perturbation theory of a conservative system,
$M-1$ of the basis vectors are
{\em not}\/ eigenvectors. Concluding remarks are given
in Section \ref{sect:concl}.

The overall results of this paper are then (a) nontrivial
examples of the Jordan block problem
in some {\em exact}\/ treatments of dissipative systems,
(b) an extended BB formalism for dealing with such situations, and (c)
a matrix representation of such Jordan blocks including their
perturbation theory.

\section{Waves in open systems}
\label{sect:waves}

\subsection{Wave equation}

We consider waves in 1 d described by
\beq
 \left[ \rho(x) \partial_t^2-\partial_x^2 \right] \phi (x,t) = 0
\eeql{eq:we}
on the half line $[0, \infty)$, with $\phi(x{=}0,t)=0$ and $\phi(x,t)$
approaching zero rapidly as $x \rightarrow \infty$ \cite{fn2}.  Let the
system $S$ be the interval $[0,a]$, and the bath $B$ be $(a, \infty )$,
with $\rho(x{>}a) = 1$.  Energy is exchanged between $S$ and $B$ only
through the boundary $x=a$.  The outgoing-wave condition $\partial
_t{\phi }(x,t)=-\partial _x \phi (x,t)$ is imposed for $x>a$.

The physical relevance of this model in describing
strings \cite{string},
electromagnetism \cite{lang} and
gravitational waves \cite{price}, as well as the mathematical formalism,
have been discussed elsewhere \cite{bior,comp1,comp2,tong,RMP}.

Everywhere below we restrict to $\rho(x)>0$, which is necessary both
physically ($\rho$ is a density \cite{string,price} or a dielectric
constant \cite{lang}) and mathematically (the equation becomes singular
if $\rho(x_0)=0$ for some $x_0$).  In particular, this means that
$\rho(x)$ can contain a (positive) $\delta$-function, but not $\delta'$
or higher derivatives (for which equations such as (\ref{eq:eigenwe})
would be distributionally undefined anyway).  Thus, $\phi$ has to be
continuous, but $\phi'$ may have discontinuities.

The eigenfunctions or {\em quasinormal modes\/} (QNMs) are factorized
solutions
\beq
  \phi(x,t) = f_j(x) e^{-i\omega_j t}\;.
\eeql{eq:fact}
The spatial function satisfies
\beq 
  [\partial _x^2 + \rho (x) \omega _j^2] f_j(x) = 0\;,
\eeql{eq:eigenwe}
with $f_j(x) \propto e^{i\omega_j x}$ for $x>a$.  Introduce the
conjugate momentum ${\hat \phi} = \rho(x) \partial_t \phi$, and the
two-component vector
\beq
  \ket{\phi} = \pmatrix{ \phi \cr {\hat \phi } }\;.
\eeql{eq:twocompdef}
In terms of this, the dynamics can be cast into the
Schr\"odinger equation $i\partial_t\ket{\phi}=H\ket{\phi}$ with the NHH
\beq
  H = i \pmatrix{0 & \rho (x)^{-1} \cr  \partial _x^2 & 0}\;.
\eeql{eq:hwe}
The identification ${\hat \phi}=\rho \partial_t \phi$ follows from the
evolution equation \cite{bj}. In this two-component form, the
eigenvectors are
\beq
  \ket{f_j} \equiv \pmatrix{f_j \cr {\hat f}_j } =
  \pmatrix{f_j \cr -i\omega_j \rho f_j }\;.
\eeql{eq:eigentwocomp}
We shall define the duality transformation only after the dynamics
has been discussed.

The hamiltonian (\ref{eq:hwe}) is non-hermitian, as will be obvious
once the appropriate inner product (\ref{eq:inner2}) is defined.
Nevertheless, on the ``universe'' $[0,\Lambda]$ (with a nodal condition
at $x=\Lambda\rightarrow\infty$) a complete real spectrum of ``universe
modes'' is guaranteed to exist. The latter are given by
(\ref{eq:eigentwocomp}) for $\pm\omega_j$, where $f_j$ are the
eigenfunctions with eigenvalue $\omega_j^2$ of the operator
$-\rho(x)^{-1}\partial_x^2$, which is hermitian and positive on the
``one-component'' space with inner product $\langle
u,v\rangle=\int_0^\Lambda\!dx\,\rho(x)u^*(x)v(x)$. This indirect
construction of the spectrum of (\ref{eq:hwe}) fails on the interval
$[0,a]$ (the energy of which is not conserved), however, since the very
definition of the outgoing-wave condition $\hat{\phi}(a^+)=-\phi'(a^+)$
is not possible with only one component. Consequently, the QNM
frequencies will have a nonvanishing $\im\omega_j<0$.

\subsection{Green's function and poles}

The dynamics of these open wave systems is best discussed
in terms of the Green's function $G(x,y;t)$, viz.,
\beq
 \phi(x,t) = \int_0^{\infty}  \left[
 G(x,y;t) {\hat \phi}(y) + \partial_t G(x,y;t) \rho(y) \phi(y) \right] dy\;,
\eeql{eq:greenev}
where $\phi$ and ${\hat \phi}$ are the initial values of the
wavefunction and conjugate momentum.  The behavior of $G$ is completely
given by the contributions from the singularities in the frequency
domain of its Fourier transform ${\tilde G}(x,y;\omega)$.  It can be
shown \cite{comp1} that under two conditions, viz.,
(a) $\rho(x)$ has a discontinuity at $x=a$ to provide a natural
demarcation of the system from its surroundings (discontinuity
condition), and 
(b) $\rho(x) = 1$ for $x>a$,
so that outgoing waves are not scattered back into
the system (no-tail condition), the only contributions will be from 
the isolated poles associated with the
eigenvectors, and {\em if all poles are simple\/} then $G$ can be
represented as
\beq
  G(x,y;t) = i \sum_j \frac{f_j(x) f_j(y)}{ (f_j, f_j) } e^{-i\omega_j t}\;.
\eeql{eq:grep}
The normalization factor $(f_j , f_j)$, which will turn out to be the
same as $\prod{f^j}{f_j}$ and therefore central to the issue at hand,
will be derived and discussed in detail below. Note that the numerator
of $G$ goes as $f_j(x) f_j(y)$, not as, e.g.,
$f_j^{\vphantom{*}}(x)f_j^*(y)$ (which indeed would violate the
symmetry of $G$ in $x$ and $y$). Thus $(f_j, f_j)$ will have to go as
$f_j^2$, not $|f_j^2|$; this makes a crucial difference, since QNM
wavefunctions in general are not real.

The derivation of the pole contributions is given in \cite{comp1}, and
the main elements will be reviewed below as preparation for Section
\ref{sect:dual}.  One starts from the defining equation for the
frequency-domain Green's function
\bea
  D(\omega) \, {\tilde G}(x,y;\omega)
  &\equiv &\left[ \partial_x^2 + \rho(x) \omega^2 \right]
  {\tilde G}(x,y;\omega) \nonumber \\
  &=& -\delta(x-y)\;.
\eeal{eq:gomdef}
The solution for $0 \le x \le y \le a$ is given explicitly by
\beq
  {\tilde G}(x,y;\omega) = \frac{f(x,\omega) g(y,\omega) }{W(\omega)}\;,
\eeql{eq:gomsol}
where $f$ and $g$ are solutions to the homogeneous equation $D(\omega)
f = D(\omega) g = 0$, with $f$ satisfying the left boundary condition
$f(x{=}0,\omega) =0$, and $g$ satisfying the right boundary condition
$g(x,\omega) \propto e^{i\omega x}$ for $x>a$. Their wronskian is
$W(\omega) = f'g - fg'$, so that the combination (\ref{eq:gomsol}) is
independent of the normalization of $f$ and $g$.

One next writes an inverse Fourier integral to obtain $G$ in the time
domain, and distorts the integral from the real $\omega$ axis to a
large semicircle in the lower half-plane.  Under the conditions stated,
the contribution from the large semicircle vanishes and there are no
cuts (or other singularities) associated with the tail of $\rho$
\cite{comp1,comp2}; one is then left with the residues at the poles,
namely the zeros of $W$ \cite{simple}.  (If the conditions do not hold,
there will be non-pole contributions from the large semicircle and/or
cuts, and the discrete eigenvectors will be incomplete in the first
sense mentioned in Section \ref{sect:intro}, leading to transients for
short times and power-law tails for long times \cite{tail}.  This
scenario will not be pursued further here.)

At a zero $\omega_j$ of $W$, the functions $f$ and $g$ are linearly
dependent: $f_j(x) \equiv f(x,\omega_j) = C_j g(x,\omega_j)$.  Thus
$f_j$ satisfies {\em both}\/ the left and right boundary conditions,
and is an eigenfunction.  Assuming for the moment that $\omega_j$ is a
simple zero, the residue is related to $d W(\omega_j) / d\omega$, and
it is straightforward to show that
\bea
  -C_j \frac{d W(\omega_j)}{d\omega}
  &=& 2\omega_j \int_0^{a^+} \rho(x) f_j(x)^2\,dx + if_j(a)^2 \nonumber \\
  &\equiv& (f_j , f_j)\;,
\eeal{eq:normdef1}
while in the numerator we have $f(x,\omega)g(y,\omega) =
C_j^{-1}f_j(x)f_j(y)$.  The representation (\ref{eq:grep}) for the
Green's function then follows trivially. In particular, the calculation
shows why $(f_j, f_j)\propto f_j^2$ instead of $|f_j^2|$. The
definition (\ref{eq:normdef1}) of $(f_j, f_j)$ is valid for
eigenfunctions only, and will be shown to be a special case of
(\ref{eq:inner1}).

This derivation and especially (\ref{eq:normdef1}) make it clear that
the orthonormalization problem is the same as the possibility of
higher-order zeros in $W$, i.e., of non-trivial Jordan blocks, as was
already apparent in the general analysis of Section~\ref{jb-intro}.

\subsection{Duality transformation}

While the duality transformation is simple and has been presented
elsewhere \cite{bior}, it is useful to emphasize how it comes about
naturally from the dynamics.  For this purpose, we put (\ref{eq:grep})
into (\ref{eq:greenev}) and find that $\ket{\phi(t)}$
(cf.\ (\ref{eq:twocompdef})) is given by (\ref{eq:expt}), with the
coefficients~\cite{spanning}
\beq
 a_j = \frac{i}{ (f_j, f_j) } \left\{ 
 \int_0^{a+} \left[ f_j(y) {\hat \phi(y)} + {\hat f_j}(y) \phi(y)
 \right] \, dy  + f_j(a) \phi(a)  \right\}.
\eeql{eq:proj2}
All reference to initial data on the ``outside" $x>a$ has been
eliminated.  Physically, the outgoing condition ensures that the
outside data do not propagate in; mathematically, both the initial data
and the Green's function are annihilated on the ``outside" by
$\partial_t + \partial_x$.

The projection formula (\ref{eq:proj2}) suggests the definition of a bilinear
map \cite{fnnotation} between two vectors
\beq
  ( \psi , \chi ) =  i\left[\int_0^{a^+}
  (\psi {\hat \chi} + {\hat \psi} \chi )\, dx+\psi(a) \chi(a) \right]\;.
\eeql{eq:inner1}
It is seen that the definition (\ref{eq:inner1}) is consistent with
(\ref{eq:normdef1}).  Using this notation, we can now write the
projection formula (\ref{eq:proj2}) compactly as
\beq
  a_j = \frac{ (f_j , \phi) }{ (f_j , f_j )}\;.
\eeql{eq:projcomp}

The map (\ref{eq:inner1}) has two important properties:
(a) in the integral it cross-multiplies the two components, and
(b) there is no complex conjugation of the first argument.
These suggest the definition of a duality transformation ${\cal
D}={\cal F}$, with ${\cal F}$ a flip map that (a) interchanges the two
components and (b) complex conjugates (cf., e.g., \cite{mishra}), viz.,
\beq
  {\cal F}  \pmatrix{ \psi_1 \cr \psi_2} \equiv
  -i  \pmatrix{ \psi_2^* \cr \psi_1^* }\;,
\eeql{eq:ddef}
in terms of which the bilinear map can be related to the standard inner
product between two-component vectors
\beq
( \psi , \chi ) = \prod{{\cal F} \psi}{\chi}\;,
\eeql{eq:drelate}
where the latter is defined as
\beq
  \prod{\zeta}{\chi} \equiv
  \int_0^{\infty} ( \zeta^* \chi + {\hat \zeta}^* {\hat \chi} ) \, dx\;.
\eeql{eq:inner2}
In showing the equivalence (\ref{eq:drelate}), one has to use the
outgoing property to collapse the integral over $(a,\infty)$ in
(\ref{eq:inner2}) to the surface term~\cite{tong}. That is,
(\ref{eq:drelate}) only holds when either $\ket{\chi}$ and ${\cal
F}\ket{\zeta}$, or ${\cal F}\ket{\chi}$ and $\ket{\zeta}$, are both
outgoing, and only in these cases we will use (\ref{eq:inner2}) in this
paper. Thus, $\ket{\chi}$ and $\ket{\zeta}$ in general belong to
different spaces, and hence the structure is not precisely that of a
Hilbert space. Since this latter issue is unrelated to the possibility
of having higher-order poles, however, these functional-analytic
details will be dealt with separately \cite{math}.

Two symbols ${\cal D}$ and ${\cal F}$ have been used in anticipation of
their inequivalence in the case of higher-order poles. However, in the
general case ${\cal F}$ remains defined by~(\ref{eq:ddef});
cf.~(\ref{defD-gen}).

The property (\ref{eq:dhhd}) is equivalent to the symmetry of $H$ under
the bilinear map:
\beq
  (\psi, H \chi) = (H\psi, \chi)\;.
\eeql{eq:symh}
To verify (\ref{eq:symh}), one needs to integrate by parts; the surface
term generated exactly cancels the surface term in (\ref{eq:inner1}).
The ``orthogonality'' of QNMs,
\beq
  (f_j,f_k)=0\;,\qquad j\neq k
\eeql{ortho}
now follows in an immediate transcription of the standard proof.

What remains then is to show that higher-order zeros in $W$ can indeed
exist (Section~\ref{sect:hp}) and to generalize the above formalism to
those cases (Sections \ref{sect:jb}--\ref{sect:degpt}).

\subsection{Klein--Gordon equation}

Also of interest in this context is
the Klein--Gordon equation

\beq
\left[  \partial_t^2-
\partial_x^2  + V(x) \right] \phi (x,t) = 0\;.
\eeql{eq:kg}

\nid
Among other things, this equation describes the propagation
of linearized gravitational waves on the curved background
of a black hole \cite{chand}.  

Essentially the entire formalism for the wave equation applies,
{\em mutatis mutandis\/} \cite{comp2}.   First, the discontinuity
condition now refers to $V(x)$, and the no-tail condition 
requires $V(x)=0$ for $x>a$.  Everywhere else 
we replace $\rho(x) \mapsto 1$, 
$-\partial_x^2 \mapsto -\partial_x^2 + V(x)$.
An example will be given below in terms of this equation.

\section{Higher-order poles}
\label{sect:hp}

\subsection{General remarks}
\label{hp-gen}

For 1-d {\em conservative}\/ systems, e.g., (\ref{eq:eigenwe}) with
nodal boundary conditions at $x=0$ and $x=a$, $W$ can only have simple
zeros since in (\ref{eq:normdef1}) the surface term is now absent,
while the integral is positive definite up to an overall phase. Thus,
the eigenfrequencies of conservative systems have a finite spacing
$\Delta\omega$, and can be labeled by the number of nodes of the
corresponding eigenfunctions. Hence, it is by no means obvious that
higher-order poles can exist in the case of outgoing waves. For this
reason, we will first demonstrate some examples of higher-order poles
(cf.~\cite{bell}) before studying the extension of the BB formalism.
For simplicity, here we shall concentrate on second-order poles. The
demonstration of the existence of third-order poles and the discussion
of other possibilities are deferred to Appendix~\ref{higher}.

Second-order poles can be obtained by allowing two first-order poles to
merge.  The converse is also true: any second-order pole, when suitably
perturbed, splits into two first-order ones, as demonstrated explicitly
in Section~\ref{sect:degpt}. One is thus permitted to think generically
of the coalescence of poles, and this leads to the concept of the
``loss" of eigenvectors.  Consider two nearby first-order zeros of $W$,
associated with two distinct eigenfunctions, say $f_j$ and $f_k$. But
when the poles merge, there is only {\em one\/} eigenfunction---in
contrast to the case of degeneracies in conservative systems.  To see
this, it suffices to notice that for a given $\omega$, the boundary
conditions $f(x{=}0,\omega)=0$ and $f'(x{=}0,\omega)=1$, say, uniquely
specify one function.  The ``lost" or ``missing" eigenfunction makes
the possibility of higher-order poles all the more interesting.

It is best to look for second-order poles on the imaginary axis in the
$\omega$ plane.  First, apart from some overall factors of $i$, the
problem is purely real and easy to handle.  More physically, as the
system parameter(s) are tuned, it is ``unlikely" that two poles in the
complex plane would collide---not only would this require the
simultaneous tuning of two parameters, but one also expects level
repulsion \cite{fnrepul}.  However, eigenvectors  of the dissipative
system~(\ref{eq:we}) exist in pairs, with frequencies $\omega$ and
$-\omega^*$ lying on the same horizontal line.  It would require the
tuning of only {\em one}\/ parameter to make them collide; when they
collide, they must do so on the imaginary axis.  In fact, we expect
that after they collide, the two poles will move apart along the
imaginary axis, in exact analogy to an oscillator going through
critical damping.  This scenario is exemplified in both models shown
below, and it remains an open question whether higher-order poles can
exist off the imaginary axis.
 
\subsection{Example in the wave equation}
\label{hp-WE}

With these remarks, we now look for a double zero of $W$ in the case of
the wave equation for $\omega = -i\gamma$, with $\gamma$ positive.  The
differential equation (suppressing the mode index $j$ \cite{j0}) then
becomes real:
\beq
  \left[ \partial_x^2 - \rho(x) \gamma^2 \right] f(x) = 0
\eeql{eq:eigenim}
and the eigenvalue condition is $f'/f = \gamma$ at $x=a^+$, which
ensures that $W(-i\gamma)=0$.  For $\omega=-i\gamma$ to be a double
zero, we also need $(f , f) \propto dW/d\omega$ to vanish:
\beq
  i(f,f) = 2\gamma \int_0^{a^+} \rho(x) f(x)^2 dx- f(a)^2 = 0\;.
\eeql{eq:zeronorm1}
By using (\ref{eq:eigenim}) to express $\rho(x) f(x)$ in terms of $f''$
and then integrating by parts, this condition can be recast as
\beq
  i(f,f) = -\frac{2}{\gamma} \int_0^a f'^2 dx + f(a)^2 = 0\;.
\eeql{eq:zeronorm2}
The last term has just been reversed on account of the surface term
$-2f(a)^2$ produced in the integration by parts.

Interestingly, $\rho(x)$ does not appear in (\ref{eq:zeronorm2}), and
this is central to the construction of examples, as follows.
(a)  Choose any function $f(x)$ satisfying $f(x{=}0)=0$, $f'(x{=}0)>0$
and $f''(x) >0$.
(b) Use (\ref{eq:zeronorm2}) to determine $\gamma$.
(c)  Put these back into (\ref{eq:eigenim}) to find $\rho(x)$,
which is guaranteed to be positive.

There is however one further subtlety.  Such a construction gives
$f'(a^-)$, and also $f'(a^+) = \gamma f(a)$; the difference between
these two must be attributed, through (\ref{eq:eigenim}), to $\rho(x) =
\cdots + \mu\delta(x-a)$, with $\mu= \gamma^{-1} -
f'(a^-)/(\gamma^2f(a))$.  One must check that $\mu\ge0$, i.e., that
\beq
  2\int_0^af'^2dx\ge f(a)f'(a^-)\;.
\eeql{eq:deltapos}
This condition is nontrivial, and for instance violated for
$f(x)=x+\alpha x^n$ and for some $\alpha$ if $n\ge5$.

Yet examples abound, e.g.,
\begin{mathletters}
\bea
  f(x<1) &=& \sinh(Kx)\label{ex1-f} \\
  \gamma &=& K\cotanh K+\frac{K^2}{\sinh^2K}\label{ex1-gamma} \\
  \rho(x)&=& \frac{K^2}{\gamma^2}\theta(1-x)
             +\frac{K^2}{\gamma^2\sinh^2K}\delta(x-1)+\theta(x-1)\;,
\eeal{eq:ex1}
\label{ex-tot}\end{mathletters}%
for any $K>0$. Note that in this example always
$\rho(0{<}x{<}1)=K^2/\gamma^2<1$, the case in which there is a
zero-mode (i.e., a single simple pole on the negative imaginary
$\omega$ axis) even if the $\delta$-term in (\ref{eq:ex1}) for $\rho$
is absent~\cite{tong,sq}. Incidentally, this case without the
$\delta$-term thus already shows that the open string model
(\ref{eq:we}) can exhibit features not found in damped harmonic
oscillators, as will become even clearer in Appendix~\ref{higher}.

\subsection{Example in the Klein--Gordon equation}

We next give an example for the Klein--Gordon equation; this example
may appear more natural in that the system (i.e., $V(x)$) is specified
in advance and not obtained as an answer.

Recall the P\"oschl--Teller potential \cite{pt}
\beq
  V(x) = V_0 \mathop{\rm sech}\nolimits^2 x\;.
\eeql{eq:pt}
The Klein--Gordon equation (\ref{eq:kg}) with this potential is exactly
soluble \cite{pt}, with the eigenvalues given by \cite{fnevenodd}
\beq
  \omega_j = \left\{ \begin{array}{ll}
  \pm \sqrt{V_0 - \quarter} - i(j+\half) & \qquad\mbox{if $V_0 \ge \quarter$} \\
  -i\left[j+\half\pm\sqrt{\quarter-V_0}\right]&\qquad\mbox{if $V_0\le\quarter$}
  \end{array} \right.
\eeql{eq:pteigen}
$j = 0, 1, 2, \dots$\ . Each pair of poles merge at the parameter value
$V_0 = \quarter$. Again, the merging is exactly like an oscillator
going through critical damping.

This example may however be regarded as slightly unsatisfactory in one
way: the potential has no discontinuity but does have a tail (i.e.,
$V(x)$ does not vanish outside a finite interval $|x|\le a$), so the
set of discrete eigenvectors would not be complete even when all the
poles are first-order.  Nevertheless, we may consider a minor
alteration: if $V(x)$ is truncated at $|x| = L$ for some large $L$, one
would expect the eigenvalues not to be much affected.  Actually this is
not the case in general~\cite{tam}, but here we shall not go into this
subtlety, which is not related to the present issue.  It is sufficient
for the present purpose that at least the {\em first\/} pair of modes
$j=0$ are not too much affected by truncation, in the sense that the $L
\rightarrow \infty$ limit recovers the pole position for the
untruncated $V(x)$.  This pair of poles at critical damping then
demonstrates a double zero of $W$ in a context where the eigenvectors
are otherwise complete.  We have verified numerically the existence of
this double pole for $L=5$, $V_0 = 0.252279109\dots$, $i\omega=
0.511109\dots$\ .

\section{Jordan blocks}
\label{sect:jb}

In this Section, we focus on a single $M$th-order zero $\omega_j$ of
$W$ and the subspace associated with it. As already mentioned in
Section~\ref{hp-gen}, in this subspace there is only one eigenvector,
and presently $M-1$ other basis vectors will be obtained. This means
that $H$ is not diagonalizable, and we consider this $M \times M$ block
in $H$. In this and the following Section, the discussion will be
confined to the wave equation (\ref{eq:we}).

Using the definitions below (\ref{eq:gomsol}), the
(position-independent) wronskian can be written as
$W(\omega)=f'(0,\omega)g(0,\omega)$. Differentiating with respect to
$\omega$, one now proves by induction that
$\partial_\omega^ng(0,\omega)|_{\omega=\omega_j}=0$ for $0\le n\le
M-1$. This means that, up to this order in $\omega-\omega_j$, the
functions $f$ and $g$ satisfy the same boundary conditions and hence
can be normalized to be equal, i.e., if we define
\beq
  f(x,\omega)=\sum_{n=0}^{M-1}f_{j,n}(x)(\omega-\omega_j)^n
  +{\cal O}[(\omega-\omega_j)^M]\;,
\eeql{taylor-f}
so that
$f_{j,n}(x)\equiv(1/n!)\partial_\omega^nf(x,\omega)|_{\omega=\omega_j}$,
then we also have
\beq
  g(x,\omega)=\sum_{n=0}^{M-1}f_{j,n}(x)(\omega-\omega_j)^n
  +{\cal O}[(\omega-\omega_j)^M]\;.
\eeql{g-f}
Now define the time-dependent functions
\bea
  f_{j,n}(x,t)&\equiv&\frac{1}{n!}\partial_\omega^n[f(x,\omega)e^{-i\omega t}
    ]_{\omega=\omega_j}\nonumber\\
  &=&\frac{1}{n!}\partial_\omega^n[g(x,\omega)e^{-i\omega t}
    ]_{\omega=\omega_j}\nonumber\\
  &=&\sum_{m=0}^{n}f_{j,n-m}(x)\frac{(-it)^m}{m!}e^{-i\omega_jt}
\eeal{def-fnt}
for $0\le n\le M-1$, where the last line follows using (\ref{g-f}).
These are not only outgoing solutions of the wave equation (since
$g(x,\omega)e^{-i\omega t}$ is such a solution for any $\omega$), but
satisfy the nodal condition at the origin as well (since
$f(x,\omega)e^{-i\omega t}$ has this property for any $\omega$). The
momenta associated with the $f_{j,n}$ are
\bea
  \hat{f}_{j,n}(x,t)&\equiv&\rho(x)\dot{f}_{j,n}(x,t)\nonumber\\
  &=&-i\rho(x)[\omega_jf_{j,n}(x,t)+f_{j,n-1}(x,t)]\;,
\eeal{fjnhat}
so that the action of the hamiltonian is
$H\ket{f_{j,n}}=\omega_j\ket{f_{j,n}}+\ket{f_{j,n-1}}$, with
$\ket{f_{j,-1}}\equiv0$.

For fixed $j$, the functions $f_{j,n}(x,t)$ ($n=0,\dots,M-1$) are
linearly independent (in the sense that a non-trivial superposition
cannot vanish identically in $x$ and $t$), as is obvious by looking at
the highest power of $t$ in each (the coefficient of
$t^ne^{-i\omega_jt}$ in $f_{j,n}$ is $\propto f_j(x)$, which by
definition does not vanish identically). Therefore the initial data
$\ket{f_{j,n}}$ have to be independent as well~\cite{vary-cst},
otherwise one would have a vanishing superposition evolving into a
nonvanishing function.

Thus the set $\{\ket{f_{j,n}}\}_{n=0}^{M-1}$ is a basis, in which the
hamiltonian reads
\beq
  H=\pmatrix{\omega_j &        1 &        0 & \cdots &                  0 \cr
                    0 & \omega_j &        1 & \cdots & \vphantom{\ddots}0 \cr
                    0 &        0 & \omega_j & \ddots &                  0 \cr
               \vdots &   \vdots &   \vdots & \ddots &             \vdots \cr
                    0 &        0 &        0 & \cdots &           \omega_j }\;.
\eeql{Hmat}
Each entry in (\ref{Hmat}) is really a $2\times2$ matrix
(cf.\ (\ref{eq:hwe})) operating on a two-component vector such as
(\ref{eq:twocompdef}), giving $H$ the structure of a $2M\times2M$
tensor product. However, the very fact that
$\{\ket{f_{j,n}}\}_{n=0}^{M-1}$ is a basis for $H$ means that each of
these $2\times2$ matrices is $\propto\openone$.

While it is guaranteed that $H$ can be cast into the so-called
Jordan-block form (\ref{Hmat}) in a subspace with precisely one
eigenvector (e.g., \cite{I&P}), we have now established a basis with
respect to which this is indeed the case, and related this basis to the
solutions $f(x,\omega)$, $g(x,\omega)$. The basis is not unique, since
a rescaling $f(x,\omega)\mapsto{\cal N}(\omega)f(x,\omega)$ mixes the
$f_{j,n}$, with only $f_j=f_{j,0}$ remaining invariant up to a
prefactor; in fact, this rescaling is readily checked to generate
precisely those basis transformations which leave the form (\ref{Hmat})
for $H$ invariant. A further---essentially unique---specification of
the basis will be made in Section~\ref{sect:dual}.

Vectors from different blocks are ``orthogonal'' under the bilinear map
(\ref{eq:inner1}), i.e.,
\beq
  (f_{j,n},f_{k,m})=0\;,\qquad j\neq k\;.
\eeql{blk-ortho}
The proof proceeds by induction with respect to $n+m$. The case $n+m=0$
is the standard one of eigenvectors given in (\ref{ortho}). Now
consider
\bea
  \omega_j(f_{j,n},f_{k,m})+(f_{j,n-1},f_{k,m})&=&
  (Hf_{j,n},f_{k,m})\nonumber\\
  &=&(f_{j,n},Hf_{k,m})\nonumber\\
  &=&\omega_k(f_{j,n},f_{k,m})+(f_{j,n},f_{k,m-1})\;.
\eea
On both sides, the second terms vanish by the induction hypothesis, so
one is left with $(\omega_j-\omega_k)(f_{j,n},f_{k,m})=0$, proving
(\ref{blk-ortho}).

\section{Generalized duality transformation}
\label{sect:dual}

Having obtained the extra non-eigenvector solutions $f_{j,n}(x,t)$
($n\ge1$) in the previous Section, we now investigate how these enter
into the field expansion. In doing so we shall consider all the poles
simultaneously, so that the block size $M$ acquires an index $j$. As
with the Jordan-block form of $H$ (cf.\ the remark below (\ref{Hmat})),
a basis dual to $\{\ket{f_{j,n}}\}_{n=0}^{M_j-1}$ is known to exist on
general grounds (see Appendix~\ref{dualapp} for details), but it
remains to find its explicit form and, if possible, to choose the
original basis (which still is subject to the freedom pointed out below
(\ref{Hmat})) so that the ensuing expressions will be as simple as
possible.

Our starting point is (\ref{eq:greenev}), with
$G(x,y;t)=\int(d\omega/2\pi)\hp\tilde{G}(x,y;\omega)\hp e^{-i\omega
t}$, where $\tilde{G}$ is given by (\ref{eq:gomsol}). The wronskian has
an $M_j$th-order zero~\cite{simple}
\bea
  W(\omega)&=&W_{j,M_j}(\omega-\omega_j)^{M_j}{\cal M}(\omega)\nonumber\\
  &=&W_{j,M_j}(\omega-\omega_j)^{M_j}+{\cal O}[(\omega-\omega_j)^{M_j+1}]\;,
\eeal{Wexp1}
with ${\cal M}(\omega_j)=1$. We now make use of the remaining freedom
$f(x,\omega)\mapsto{\cal N}(\omega)f(x,\omega)$, taking ${\cal
N}(\omega)={\cal M}(\omega)^{-1/2}+{\cal O}[(\omega-\omega_j)^{M_j}]$
(which is analytic in a neighborhood of $\omega=\omega_j$) and
similarly for $g$, preserving (\ref{g-f}). After this transformation we
have~\cite{normW}
\beq
  W(\omega)=W_{j,M_j}(\omega-\omega_j)^{M_j}
    +{\cal O}[(\omega-\omega_j)^{2M_j}]\;,
\eeql{Wexp}
where we draw attention to the order of the error term.
Eq.~(\ref{Wexp}) will greatly simplify the formulas below. The contour
integral for $G$ is now straightforward, leading to
\beq
  G(x,y;t)=\sum_j\frac{e^{-i\omega_jt}}{iW_{j,M_j}}
    \sum_{n=0}^{M_j-1}\sum_{m=0}^n
    f_{j,m}(y)f_{j,n-m}(x)
    \frac{(-it)^{M_j-1-n}}{(M_j-1-n)!}
\eeql{G-int}
for $t\ge0$, which we assume throughout this Section. By symmetry, (\ref{G-int}) also holds for $0\le y<x\le a$ even though this is
not the case for the original (\ref{eq:gomsol}). If $M_j=1$ for all~$j$
this agrees with the known result for simple poles, which follows by
combining (\ref{eq:grep}) and (\ref{eq:normdef1}). In general, one can
rewrite
\beq
  G(x,y;t)=\sum_j\frac{1}{iW_{j,M_j}}
    \sum_{n=0}^{M_j-1}f_{j,M_j-1-n}(y)f_{j,n}(x,t)
\eeq
in terms of the functions $f_{j,n}(x,t)$ defined in (\ref{def-fnt}).
Insertion into (\ref{eq:greenev}) yields the time evolution
\beq
  \phi(x,t)=-\sum_j\frac{1}{W_{j,M_j}}\sum_{n=0}^{M_j-1}
  \left(f_{j,M_j-1-n}\,,\,\phi\right)f_{j,n}(x,t)
\eeql{blk-exp}
in terms of the bilinear map (\ref{eq:inner1}). In particular, this
holds for $\phi=f_{k,m}$, in which case terms with $j\neq k$ vanish by
(\ref{blk-ortho}), and the linear independence of the $f_{k,n}(t)$
discussed above (\ref{Hmat}) implies that the coefficients on both
sides of (\ref{blk-exp}) are equal, i.e.,
\beq
  \left(f_{j,n},f_{k,m}\right)=
    -\delta_{jk}\delta_{n+m,M_j-1}W_{j,M_j}\;.
\eeql{bi-ortho}
Of course, this intra-block ``orthogonality'' relation is conditional
on the normalization $W^{(n)}(\omega_j)=0$ for $M_j+1\le n\le 2M_j-1$,
imposed in (\ref{Wexp}); without this normalization one has (\ref{fgW})
below instead. In view of (\ref{eq:drelate}), the relation (\ref{bi-ortho})
leads one to define
\beq
  \ket{f^{j,n}}\equiv{\cal D}\ket{f_{j,n}}\equiv{\cal F}\ket{f_{j,M_j-1-n}}\;,
\eeql{defD-gen}
where ${\cal F}$ is the flip operation in (\ref{eq:ddef}). Thus
\beq
  \prod{f^{j,n}}{f_{j'\!,n'}}=-W_{j,M_j}\delta_{jj'}\delta_{nn'}\;,
\eeql{blk-dual}
where the constant of proportionality $W_{j,M_j}$ is nonzero by
definition (cf.\ (\ref{Wexp1})), so that there is no normalization
problem. Equation (\ref{blk-dual}) is a significant result; it shows
that, unless $M_j=1$ for all $j$, the duality map ${\cal D}$ no longer
coincides with ${\cal F}$: ${\cal D}$ changes the intra-block index $n$
of $\ket{f_{j,n}}$. Since it is the flip map which obeys
\beq
  {\cal F}H=H^{\dagger}{\cal F}
\eeql{fhhf}
(note that the proof outlined below (\ref{eq:symh}) does not invoke any
assumptions on the block structure of $H$), the relation
(\ref{eq:dhhd}) in general is {\em not\/} satisfied by the operator
${\cal D}$ implicit in~(\ref{eq:comp1}). Since it is an immediate
consequence of (\ref{fhhf}) that $\cal F$ carries right into left
eigenvectors and vice versa, the left eigenvector corresponding to
$\ket{f_j}=\ket{f_{j,0}}$ is $\bra{{\cal
F}f_{j,0}}=\bra{f^{j,M_j-1}}\neq\bra{f^{j,0}}$ for $M_j>1$. While the
left and right eigenvectors thus are orthogonal as stipulated in
Section~\ref{jb-intro}, this does not lead to orthonormalization
problems as these vectors are not the dual of each other.

Using (\ref{bi-ortho}) and (\ref{defD-gen}), one may write the final
result for the generalized bi-orthogonal expansion as
\beq
  G(x,y;t)=i\sum_j\sum_{n=0}^{M_j-1}
    \frac{f_{j,M_j-1-n}(y)f_{j,n}(x,t)}{(f_{j,M_j-1-n}\,,\,f_{j,n})}\;,
\eeql{Gexp-gen}
so that the time evolution reads
\beq
  \ket{\phi(t)}=\sum_j\sum_{n=0}^{M_j-1}
    \frac{\prod{f^{j,n}}{\phi}}{\prod{f^{j,n}}{f_{j,n}}}
    \ket{f_{j,n}(t)}\;.
\eeql{ket-exp}
The $t\downarrow0$ limit of $G(x,y;t)$ then yields the sum rule
\beq
  i\sum_j\sum_{n=0}^{M_j-1}
    \frac{f_{j,M_j-1-n}(y)}{(f_{j,M_j-1-n}\,,\,f_{j,n})}
    \pmatrix{f_{j,n}(x) \cr \hat{f}_{j,n}(x)}=\pmatrix{0 \cr \delta(x-y)}\;,
\eeql{sumr}
while in the same limit, (\ref{ket-exp}) is indeed seen to be of the
form (\ref{eq:comp1}).

Equation (\ref{bi-ortho}) is the generalization of (\ref{eq:normdef1})
and (\ref{ortho}), while our proof is a slight simplification even in
the simple-pole case, cf.~\cite{comp1}. The representation
(\ref{Gexp-gen}) generalizes (\ref{eq:grep}), and (\ref{ket-exp})
extends (\ref{eq:expt}) and (\ref{eq:projcomp}). Also the simple-pole
counterpart of (\ref{sumr}) is already known \cite{comp1}.

One may ask to what extent the basis we have obtained is unique. On the
one hand we demand that in our basis the Hamiltonian have the block
form (\ref{Hmat}), and below this equation it has already been remarked
that this forces the functions $f_{j,n}$ to be of the form
(\ref{def-fnt}) for some normalization of $g(x,\omega)$. On the other
hand, to ensure the simplicity of formulas like (\ref{ket-exp}), we
require that the dual to $\ket{f_{j,n}}$ be some ${\cal
F}\ket{f_{j',n'}}$, where (\ref{blk-ortho}) then forces $j'=j$. Now for
{\em any\/} normalization of $f$ and $g$ (i.e., temporarily abandoning
(\ref{g-f}) and (\ref{Wexp})) one has
\beq
  (f_{j,n},g_{j,m})=-W_{j,n+m+1}
\eeql{fgW}
as long as $n,m\le M_j-1$ (implying that the bilinear map vanishes if
$n+m\le M_j-2$), as can be proved by operating with
$\sum_{\ell=0}^n[(n-\ell)!(m+\ell+1)!]^{-1}\partial_{\omega}^{n-\ell}
\partial_{\omega'}^{m+\ell+1}|_{\omega=\omega'=\omega_j}$
on the identity
\beq
  (\omega^2-\omega'^2)\int_0^{a^+}\!\rho f(\omega)g(\omega')\,dx=
  [i\omega'f(a,\omega)-f'(a^+,\omega)]g(a,\omega')+f'(0,\omega)g(0,\omega')\;.
\eeql{Wident}
Hence, the product $(f_{j,n}\,,\,f_{j,M_j-1-n})$ is always
nonvanishing, and the only way of achieving bi-orthogonality by setting
other products to zero as in (\ref{bi-ortho}) is to normalize $W$ as in
(\ref{Wexp}), which obviously fixes the $\ket{f_{j,n}}$ up to one
overall prefactor per Jordan block. Hence, our choice of basis is
unique up to these prefactors. In fact, further analogy to the customary treatment\cite{RMP} of the simple-pole case results if one sets $W_{j,M_j}=-2\omega_j$ for all $j$, so that $\left(f_{j,n},f_{k,m}\right)=2\omega_j\delta_{jk}\delta_{n+m,M_j-1}$. With this preferential normalization, which is convenient in applications\cite{path}, the freedom discussed in this paragraph is reduced further to one overall sign per Jordan block.

In closing this account of the generalized duality transformation, let
us return to the example of Section~\ref{hp-WE} with $M=2$. Using
either (\ref{taylor-f}) or the integral representation~\cite{vary-cst},
one can find the ``preferred" second basis function (i.e., the one for
which $(f_{0,1},f_{0,1})=0$~\cite{j0}) corresponding to $f_0=f$ as in
(\ref{ex1-f}) as
\begin{mathletters}
\bea
  f_{0,1}(x)&=&f_a(x)+f_b(x)\;,\label{f01ex}\\
  f_a(x)&=&i\frac{K}{\gamma}x\cosh(Kx)\;,\label{fa-ex}\\
  f_b(x)&=&-i\left(\frac{2}{3}\frac{K}{\gamma}
           +\frac{1}{2K}\right)\tanh K\sinh(Kx)\;,
\eeal{fb-ex}
\label{f01tot}\end{mathletters}%
where the conjugate momenta $\hat{f}_a$ and $\hat{f}_b$ are given by
(\ref{fjnhat}). While (\ref{f01tot}) is given here for reference and
further use in Sec.~\ref{sect:degpt}, there does not seem to be a
simple physical interpretation of this result. The normalization
occurring in the field expansion (\ref{ket-exp}) is evaluated to be
\beq
  \prod{f^0}{f_0}=\prod{f^{0,1}}{f_{0,1}}=(f_0,f_{0,1})
  =\frac{K^3}{\gamma^2}\cotanh K\;.
\eeql{norm-ex}
Since the contribution of $f_b$ to the product (\ref{norm-ex}) is
$\propto(f_0,f_0)$, it vanishes. However, in a calculation in
Section~\ref{degpt-ex} this term in $f_{0,1}$ will be seen to be
essential for arriving at the correct result.

\section{Jordan-block perturbation theory}
\label{sect:degpt}

\subsection{Formalism for the generic case}
\label{degpt-gen}

The use of BBs places dissipative systems into a framework
very similar to that for conservative systems.  As a result,
it is straightforward to develop time-independent perturbation
theory, in effect by transcribing textbook results for conservative
systems, which nevertheless apply to the {\em complex}\/ eigenvalues
and eigenvalue shifts \cite{tong,fnmap}.

It will presently be investigated how this formalism, previously
developed only for QNM spectra with simple poles, is modified if at
least one eigenvalue $\omega_j$ is associated with a nontrivial Jordan
block. The ensuing splitting of the multiple pole into $M_j$ distinct
ones (in the generic case, to be defined below) is reminiscent of the
lifting of a degeneracy by a perturbation (typically breaking some
symmetry) in hermitian systems; however, important differences need to
be pointed out. In the first place, in the models here under discussion
an $M_j$th-order pole is associated with $M_j$ degrees of freedom, of
which only one is an eigenvector, as has been emphasized
before~\cite{degenerate}. Secondly, in the Jordan-block case the
splitting generically is governed by only {\em one\/} complex
parameter, and as a result the $M_j$ complex frequency shifts are not
independent; in fact, all their relative magnitudes and phases are
predetermined, and only the overall magnitude and phase depend on the
details of the perturbation, namely on the one complex parameter;
cf.~(\ref{tildeOm}). While this feature may seem unusual compared to
conservative systems, it actually simplifies the calculation.

While Jordan-block perturbations thus differ essentially from
perturbations of degenerate levels in conservative systems, we will
study the former by a method which can also be used for the latter:
transferring part of the perturbing NHH $H'$ to the unperturbed $H_0$
and treating this part exactly, upon which the remainder of $H'$ can be
dealt with using conventional non-degenerate perturbation theory.

Specifically, let $H=H_0+\lambda H'$, where $H_0$ is assumed to have a
known Jordan-block structure as described in Sections \ref{sect:jb}
and~\ref{sect:dual}, $H'$ accounts for a change in density
$\delta(\rho^{-1})$, and where $|\lambda|\ll1$. For simplicity of
notation, it is supposed that there is only one $M_j\ge2$; the
generalization to several Jordan blocks is immediate. Consider the
splitting part of $H'$:
\beq
  H_{\rm s}'\equiv
  \frac{\ket{f_{j,M_j-1}}\bra{f^{j,M_j-1}}}{\prod{f^{j,M_j-1}}{f_{j,M_j-1}}}
  H'\frac{\ket{f_j}\bra{f^j}}{\prod{f^j}{f_j}}\;.
\eeql{defHs}
This has only one nonvanishing matrix element, viz.,
\bea
  \alpha&\equiv&
  \frac{\prod{f^{j,M_j-1}}{H'|f_j}}{\prod{f^{j,M_j-1}}{f_{j,M_j-1}}}\nonumber\\
  &=&\frac{(f_j,H'f_j)}{(f_{j,M_j-1},f_j)}\label{alpha-map}\\
  &=&\frac{\omega_j^2\int_0^{a^+}\!\delta(\rho^{-1})\rho^2f_j^2\,dx}
          {(f_{j,M_j-1},f_j)}\;.
\eeal{def-alpha}
We transfer this part to the unperturbed hamiltonian:
$\tilde{H}_0\equiv   H_0+\lambda H_{\rm s}'$, so that the remaining
perturbation is $\tilde{H}'=H'-H_{\rm s}'$. The perturbation is said to
be generic iff $\alpha\neq0$. This definition of a generic perturbation
will be justified below, by showing that for sufficiently small
$\lambda$ the matrix elements of $\tilde{H}'$ effect only a
higher-order correction compared to the splitting caused by
$\alpha\neq0$. First of all, however, it should be noted that the
representation (\ref{def-alpha}) for an infinitesimal $\delta\hn\rho$
is evaluated to read $\alpha\propto\int_0^{a^+}\!\!\delta\hn\rho
f_j^2\,dx$. Using the method of variation of the constant (cf.\ the
inner integrand in~\cite{vary-cst}) and this latter representation,
$\alpha\neq0$ is seen to be equivalent to $\partial_\lambda
g(0,\omega_j)\propto\partial_\lambda W(\omega_j)\neq0$. In other words,
if $\alpha\neq0$ the $M_j$th-order zero in the wronskian at
$\omega=\omega_j$ is split already in lowest order in $\lambda$.

The eigenvalue problem for $\tilde{H}_0$ can be solved in the
$M_j\times M_j$ block associated with~$\omega_j$. For the
characteristic polynomial in this block one has
$\det(\tilde{H}_0-\omega\openone)=
(\omega_j-\omega)^{M_j}-(-)^{M_j}\lambda\alpha$,
yielding the eigenfrequencies as
\beq
  \tilde{\omega}_{j,n}=\omega_j+s\hp e^{2\pi ni/M_j}
\eeql{tildeOm}
($n=0,1,\dots,M_j-1$), where $s=\sqrt[\raisebox{1pt}{\scriptsize$M_{\hn
j}$}]{\lambda\alpha}$ is an arbitrary but fixed choice of the root.
Thus, the splittings $\tilde{\omega}_{j,n}-\omega_j$ all have the same
magnitude $\propto|s|\propto\lambda^{1/M_j}$ and are equiangular, i.e.,
their phases have constant differences. Both their magnitude and the
overall phase are determined by $\alpha$. The corresponding
eigenvectors of $\tilde{H}_0$ are
\beq
  \ket{\tilde{f}_{j,n}}=\sum_{m=0}^{M_j-1}s^me^{2\pi nmi/M_j}\ket{f_{j,m}}\;;
\eeq
since the higher-order pole has been split into first-order ones, their
duals read simply
\beq
  \ket{\tilde{f}^{j,n}}={\cal F}\ket{\tilde{f}_{j,n}}\;.
\eeq

It remains to account for $\tilde{H}'$, using conventional perturbation
theory, by evaluating its matrix in the new basis. The validity of this
procedure is not entirely trivial, since the transformation from the
basis $\{\ket{f_{j,n}}\}$ to $\{\ket{\tilde{f}_{j,n}}\}$, effected by
the matrix $P_{mn}=s^me^{2\pi nmi/M_j}$, is singular in the limit
$\lambda\rightarrow0$. For a justification, denote the matrix elements
of $\tilde{H}'$ with respect to the old basis as
$\tilde{H}'_{nm}=\prod{f^{j,n}}{\tilde{H}'|f_{j,m}}/\prod{f^{j,n}}{f_{j,n}}$,
and evaluate the inverse transformation (in fact a discrete Fourier
inversion) as $(P^{-1})_{mn}=M_j^{-1}s^{-n}e^{-2\pi nmi/M_j}$. In the
basis which diagonalizes $\tilde{H}_0$, the perturbation is then given
as
\bea
  \frac{\prod{\tilde{f}^{j,n}}{\tilde{H}'|\tilde{f}_{j,m}}}
       {\prod{\tilde{f}^{j,n}}{\tilde{f}_{j,n}}}
  &=&\sum_{k,\ell=0}^{M_j-1}(P^{-1})_{nk}\tilde{H}'_{k\ell}P_{\ell m}\nonumber\\
  &=&\sum_{k,\ell=0}^{M_j-1}
       \frac{1}{M_j}e^{2\pi(\ell m-nk)i/M_j}\tilde{H}'_{k\ell}s^{\ell-k}\;,
\eeal{transf-Hpr}
where it is crucial that the power $s^{1-M_j}$ does not occur since
$\tilde{H}'_{M_j-1,0}=0$ on account of~(\ref{defHs}). Thus the matrix
elements of $\lambda\tilde{H}'$ are ${\cal O}(s^{M_j}s^{2-M_j})$, which
means that the first-order frequency shifts due to $\tilde{H}'$ are
${\cal O}(s^2)$, small compared to the lowest-order splittings
$\Delta\tilde{\omega}_{j,n}\propto s$. With energy denominators given
by $\tilde{\omega}_{j,n}-\tilde{\omega}_{j,n'}\propto s$, higher-order
shifts are smaller still by successive powers of $s$. Of course, matrix
elements of $\tilde{H}'$ connecting the block associated with the
unperturbed $\omega_j$ to other blocks, or connecting two other blocks,
can be handled without difficulty.

\subsection{Nongeneric case}

If $\alpha=0$ the leading behaviour is determined by other matrix
elements, and the splitting of $\omega_j$ can be partial or, depending
on the scheme of calculation, occurs only in higher order~\cite{kato}.
We shall not investigate the general case, but instead give an example
of a non-generic perturbation which is relevant to the discussion in
Appendix~\ref{higher}. Namely, if a $4\times4$ Jordan block is
perturbed by the operator
\beq
  H'=\pmatrix{0 & 0 & 0 & 0 \cr \alpha & 0 &      0 & 0 \cr
              0 & 0 & 0 & 0 \cr      0 & 0 & \alpha & 0}
\eeql{Hpr4-2}
(in the basis $\{\ket{f_{j,n}}\}$), the characteristic polynomial is
found to be
$\det(H-\omega\openone)=\linebreak\relax[(\omega-\omega_j)^2-\lambda\alpha]^2$. Thus,
the fourth-order pole is split into two second-order poles, and the
latter do not undergo further splitting to any order. Of course, this
treatment does not address the question whether for some
$\delta\hn\rho$ the perturbation $H'$ can have the form (\ref{Hpr4-2})
for the open wave system (\ref{eq:we}), even if a fourth-order pole is
assumed to exist. However, $H'$ as in (\ref{Hpr4-2}) at least satisfies
the fundamental symmetry
$\prod{f^{j,n}}{H'|f_{j,m}}=\prod{f^{j,M_j-1-m}}{H'|f_{j,M_j-1-n}}$
(i.e., reflection symmetry with respect to the NE--SW diagonal within
one block), which follows from (\ref{eq:symh}) and (\ref{defD-gen}).

\subsection{Example}
\label{degpt-ex}

Returning to the example (\ref{ex-tot}) of Section~\ref{hp-WE}, and
bearing in mind the NHH action on a two-component vector given by
(\ref{eq:hwe})~\cite{rho-sing}, we study the perturbation
$\rho^{-1}(x)\mapsto\rho^{-1}(x)+\lambda\hp\theta(1-x)$.

 From its definition (\ref{alpha-map}), and using (\ref{norm-ex}) for
the normalization, one obtains
\beq
  \alpha=\frac{\lambda}{2}[K\tanh K-\sinh^2K]\;.
\eeql{alpha-ex}
That is, if $\lambda>0$ the frequency shifts $\pm\sqrt{\alpha}$ are
purely imaginary, while for $\lambda<0$ they are real and of opposite
signs. In other words, as $\lambda$ is turned from positive values
through zero to negative values, the poles move together horizontally
in the complex plane, merge, and then move apart vertically, in
accordance with the general observation made at the end of
Section~\ref{hp-gen}.

Proceeding to ${\cal O}(\lambda)$, conventional QNM perturbation theory
gives the next contributions to the shift as
$\prod{\tilde{f}^{0,0}}{\tilde{H}'|\tilde{f}_{0,0}}/
\prod{\tilde{f}^{0,0}}{\tilde{f}_{0,0}}$
and
$\prod{\tilde{f}^{0,1}}{\tilde{H}'|\tilde{f}_{0,1}}/
\prod{\tilde{f}^{0,1}}{\tilde{f}_{0,1}}$
respectively~\cite{j0}. By (\ref{transf-Hpr}), both are evaluated as
$\frac{1}{2}(H_{00}'+H_{11}')+{\cal O}(\lambda^{3/2})=H_{00}'+{\cal
O}(\lambda^{3/2})$, where the last equality follows from the symmetry
pointed out below (\ref{Hpr4-2}). In the numerator of
\beq
  H_{00}'=\frac{(f_{0,1},H'f_0)}{(f_{0,1},f_0)}\;,
\eeql{H00pr1}
the contribution of $f_b$ as in (\ref{fb-ex}) (which does not
contribute to the denominator in (\ref{H00pr1}), cf.\ below
(\ref{norm-ex})) is seen to be $\propto\alpha$ upon comparison with
(\ref{alpha-map}). For both split levels the next-order shift thus
reads
\bea
  H_{00}'&=&\frac{(f_a,H'f_0)}{(f_{0,1},f_0)}
            -i\left(\frac{2}{3}\frac{K}{\gamma}
                    +\frac{1}{2K}\right)\tanh(K)\hp\alpha\nonumber\\
         &=&i\frac{\lambda}{\gamma}\left(
              \frac{K}{4}\tanh K-\frac{K}{6}\frac{\sinh^3K}{\cosh K}
              -\frac{K^2}{4}-\frac{K^2}{12}\tanh^2K\right)
\eeal{H00pr}
and is seen to be purely imaginary, so that if the double-pole
zero-mode is split along the imaginary axis in lowest order (i.e.,
${\cal O}(\lambda^{1/2})$) the perturbed QNMs stay on this axis up to
${\cal O}(\lambda)$, again consistent with the symmetry argument of
Section~\ref{hp-gen}.

For a check, the perturbed QNMs can be also obtained directly from the
wave equation (\ref{eq:eigenwe}) together with the boundary conditions.
For $\rho^{-1}(x)$ having a constant value
$(\rho')^{-1}=\gamma^2/K^2+\lambda$ on $0<x<a$, the eigenvalue equation
is found to be
\beq
  i-\sqrt{\rho'}\cotan\left(\sqrt{\rho'}\omega\right)=
    -\omega\frac{K^2}{\gamma^2\sinh^2K}\;,
\eeql{omega-ex}
in which one has to expand
$\sqrt{\rho'}=K/\gamma-(K^3/2\gamma^3)\lambda+{\cal O}(\lambda^2)$ and
$\omega=-i\gamma+\omega_1\sqrt{\lambda}
+\omega_2\lambda+\omega_3\lambda^{3/2}+{\cal
O}(\lambda^2)$. In ${\cal O}(\lambda^0)$ and ${\cal
O}(\sqrt{\lambda})$, (\ref{omega-ex}) is satisfied identically. In
${\cal O}(\lambda)$ one obtains $(\omega_1\sqrt{\lambda})^2=\alpha$,
with $\alpha$ as in (\ref{alpha-ex}); in ${\cal O}(\lambda^{3/2})$ one
obtains $\omega_2\lambda=H_{00}'$, with $H_{00}'$ as in (\ref{H00pr}).
Hence, there is complete agreement between the Jordan-block formalism
of Section~\ref{degpt-gen} and direct expansion of the wave equation.

Finally, an example of a non-generic perturbation is furnished by
changing $K$ to $K'=K+\lambda$ in (\ref{eq:ex1}), namely by
$\delta\hn\rho(x)=\partial_K^\phs\rho(x)$, where in the differentiation
of $\rho$ its implicit $K$-dependence through $\gamma$ as in
(\ref{ex1-gamma}) must also be taken into account. In lowest order,
$H'$ shifts the double pole corresponding to $K$ to a double pole
corresponding to $K'$, and indeed $\int_0^{1^+}\!\!\delta\hn\rho
f_0^2\,dx$ is found to vanish, as stipulated below (\ref{def-alpha}).
Since beyond this leading order $H_0(K)+\lambda H'\neq H_0(K')$, the
double pole will be split eventually. In line with the treament in
Section~\ref{degpt-gen}, however, this is not pursued further here.

\section{Conclusion}
\label{sect:concl}

A remark is in place on the {\em relevance\/} of the issue considered in this paper. The above and especially Section~\ref{sect:degpt} make clear that the set of systems for which non-trivial Jordan blocks occur is of measure zero in parameter space. However, this feature is shared with, among others, stationary points in the phase space of dynamical systems, critical points in phase diagrams (note the semantic coincidence with ``critical damping"), and degeneracies in conservative quantum systems, all of which are worthy of study and are known to determine the global structure of a system's parameter space to a much greater extent than one would think at first sight. In the case of degenerate quantum levels, a further motivation is their relation to a system's physical symmetries. While the corresponding phenomenon does not yet show up on the level of this paper, further investigation reveals that two states can merge in the superpartner of a spatially symmetric open wave system\cite{susy}.

The existence questions raised in Section~\ref{sect:hp} and
Appendix~\ref{higher} are closely related to the problem of {\em
spectral inversion}. In conservative systems, the classic inversion
problem is to determine the system (e.g., $\rho(x)$ for the wave
equation or $V(x)$ for the Klein--Gordon equation) given all the real
eigenfrequencies $\omega_j$. The solution to this classic problem is
well known~\cite{invert}. The analogous problem for open systems is to
determine $\rho(x)$ or $V(x)$ from the {\em complex\/} eigenfrequencies
$\omega_j$, or, more generally, from the singularity structure of
$\tilde{G}(x,y;\omega)$ in the $\omega$-plane. If, for example,
$\tilde{G}$ is specified to have poles of order $M_j$ (say, $M_j=4$) at
$\omega_j$, does a corresponding $\rho(x)$ or $V(x)$ exist (at least
for one in a class of such singularity configurations)? Assuming the
general inversion problem for open systems (a topic for further
investigation) to be tractable, at this stage each of the following
scenarios seem conceivable.
\begin{itemize}
\item[(a)] The inversion algorithm indeed yields a $\rho(x)$ with, say,
a fourth-order pole or a pair of off-axis double poles in its
spectrum.
\item[(b)] The inversion problem turns out to have no solutions,
yielding a non-trivial proof of the non-existence of these more exotic
configurations.
\item[(c)] This particular set of singularities points to limitations
in the inversion algorithm which otherwise might have been overlooked.
\end{itemize}
Any of these possibilities would further the understanding of QNMs in
open wave systems.

Recently, for the case of simple poles we have carried out the second
quantization of the open wave system (\ref{eq:we}) using
QNMs~\cite{sq}. The QNM expansion coefficients $a_j$
(cf.\ (\ref{eq:projcomp})) emerge as the pertinent quantum degrees of
freedom, in terms of which it is possible to eliminate the ``outside''
from the equations for the cavity evolution. This relevance to the
quantum problem further motivates the study of the mode structure of
(\ref{eq:we}), and second quantization in the case for which this
structure involves nontrivial Jordan blocks indeed turns out to be possible either by Hilbert-space methods or by exactly solving the associated path integral\cite{path}.

In closing, it may be useful to place the present work into the
following context. Many wave phenomena in nature can be described by an
evolution equation $i\partial_t\ket{\phi}=H\ket{\phi}$ (see above
(\ref{eq:hwe})) and a natural question is: what are the possible forms
for $H$---leading to various types of time evolution for
$\ket{\phi}$---and how are these exemplified in physical systems? The
most familiar examples are conservative systems, for which $\ket{\phi}$
is expandable in a complete basis of normal modes, in terms of which
$H$ would be diagonal with real eigenvalues. Our earlier
work~\cite{bior,tong} shows another realization: in a large class of
outgoing wave systems, $\ket{\phi}$ is again expandable in a complete
basis (of QNMs), in terms of which $H$ would again be diagonal, but
with complex eigenvalues. The QNMs, however, are not orthogonal under
the standard inner product; therefore it is convenient to introduce
their duals as well, together with which they constitute a BB. The
present work has identified and studied a further generalization
pertaining to such open wave systems, for which $H$ is not
diagonalizable. In these circumstances we have shown that a
well-defined Jordan-block structure emerges, involving a nontrivial
duality transformation.

\section*{Acknowledgment}

This work is supported in part by the Hong Kong Research Grants Council
(Grant no. 452/95P). 

\appendix

\section{Third-order poles and beyond}
\label{higher}

As stated in Section~\ref{hp-gen}, at present it is not known whether
higher-order poles can exist off the imaginary axis. The analogy to
damped harmonic oscillators suggests that they can not, but this does
not lead to a proof directly since the harmonic oscillator picture
itself is contingent on the QNM spectrum being simple, with at most a
double-pole zero-mode. In fact, already at the end of
Section~\ref{hp-WE} it has been remarked that the analogy is imperfect,
and also intuition which is mainly based on the conservative and WKB
limits could bias one against more exotic possibilities.

A possible strategy for looking for double poles off the imaginary axis
is to first construct a fourth-order zero-mode, which upon applying a
suitable non-generic perturbation could be split into a pair of such
double poles, cf.\ Sec.~\ref{sect:degpt}. Thus returning to the case
$\omega=-i\gamma$, (\ref{fgW}) shows that for a fourth-order pole one
needs $(f_0,f_{0,1})=(f_{0,1},f_{0,1})=0$~\cite{j0}. Expanding
$f_{0,1}$ by its integral representation~\cite{vary-cst}, using
(\ref{fjnhat}) for its momentum, and eliminating $\rho$ and $\gamma$ by
(\ref{eq:eigenim}) and (\ref{eq:zeronorm2}) respectively, one arrives
at two functional equations for $f_0=f$~\cite{numint}:
\bea
  \gamma^2W_{0,2}&=&4\int_0^1\!\frac{dx}{f^2(x)}\biggl[\int_0^x\!dy\,
    f''(y)f(y)\biggr]^2-\int_0^1(f')^2\,dx=0\label{thirdO}\\
  i\gamma^3W_{0,3}&=&8\int_0^1\!dx\,f''(x)f(x)\biggl[\int_x^1\!
    \frac{dy}{f^2(y)}\int_0^y\!dz\,f''(z)f(z)\biggr]^2\nonumber\\
    &&-4\int_0^1\!\frac{dx}{f^2(x)}\biggl[\int_0^x\!dy\,f''(y)f(y)\biggr]^2=0\;,
\eeal{fourthO}
where without loss of generality we have chosen $a=1$, and where
(\ref{thirdO}) alone implies a third-order pole. Solutions to these
equations are to be sought among the functions $f$ satisfying
conditions~(a) below (\ref{eq:zeronorm2}) and the
inequality~(\ref{eq:deltapos}).

Up to now we have only investigated third-order poles. The strategy is
to seek a function $f$ satisfying (\ref{thirdO}) and, once one is
found, to construct the corresponding $\rho(x)$ using
(\ref{eq:eigenim}). We have done so within the class of functions
$f(x)=x+\alpha x^n$ ($n>2$) mentioned below
(\ref{eq:deltapos})~\cite{rho0}. It is easy to see that $W_{0,2}<0$
both for small and for large $\alpha$: (a) for small $\alpha$,
$f''\approx0$ so that the second term in (\ref{thirdO}) dominates; (b)
for large $\alpha$, one can neglect the linear term and hence find that
$\gamma^2W_{0,2}=-\alpha^2n^2(4n-3)/(2n-1)^3<0$. However, there exist
$n$ for which $W_{0,2}$ can become positive for $\alpha$ in an interval
$(\alpha_1,\alpha_2)$---so that $W_{0,2}=0$ at $\alpha_1$ and at
$\alpha_2$---for example $(\alpha_1,\alpha_2)=(2.059,3.8209)$ for
$n=5$, and $(\alpha_1,\alpha_2)=(1.063096,8.30908)$ for $n=6$. In both
cases, the inequality~(\ref{eq:deltapos}) is violated at $\alpha_1$ but
satisfied at $\alpha_2$, implying that third-order poles indeed do
exist. Besides being a stepping-stone in the seach for fourth-order
(and hence off-axis) poles, this result in itself already justifies the
general (i.e., not limited to $M_j\le2$) setup in
Sections~\ref{sect:jb}--\ref{sect:degpt}.

\section{Constructing dual bases}
\label{dualapp}

In Section~\ref{sect:dual}, one is faced with the problem of
calculating the basis dual to $\{\ket{f_{j,n}}\}_{n=0}^{M_j-1}$. There
is a standard result for finite-dimensional spaces~\cite{I&P}, which
however applies only when the dual basis is constructed {\em within\/}
the space spanned by the original one. Here we are concerned with the
original basis in $V={\cal L}[\{\ket{f_{j,n}}\}_{n=0}^{M_j-1}]$ (${\cal
L}$ denotes the linear span), but with the dual basis in a {\em
different\/} space $W$. (Guided by the simple-pole case, we expect
$W={\cal F}V$ with $\cal F$ as in (\ref{eq:ddef}), but this property
will not be used below.)

Therefore we are led to consider the following problem. Let $V$ be an
$M$-dimensional subspace of a Hilbert space with basis
$\{v_n\}_{n=0}^{M-1}$, and let $W$ be another $M$-dimensional subspace
of the same Hilbert space. Under what conditions will there be a dual
basis $\{w^n\}$ in $W$, in the sense that $\prod{w^m}{v_n}=\delta^m_n$?
We claim that the necessary and sufficient condition is
\beq
  W\cap V^\perp=\{0\}\;,
\eeql{WVperp}
where $V^\perp$ is the orthogonal complement to $V$. (For example, if
the whole Hilbert space is 3-d, and if $V$ is the $x$-$y$ plane, then
$W$ must not contain the $z$-axis.)

For a proof, let $\{\tilde{w}_n\}$ be any basis for $W$. Then the
duality of $\{w^n\}$ and $\{v_n\}$ is equivalent to
\beq
  \pmatrix{ \prod{\tilde{w}_0}{v_0} & \cdots & \prod{\tilde{w}_0}{v_{M-1}} \cr
            \vdots              &        & \vdots              \cr
            \prod{\tilde{w}_{M-1}}{v_0}&\cdots&\prod{\tilde{w}_{M-1}}{v_{M-1}}}
  \pmatrix{ w^0 \cr \vdots \cr w^{M-1} } =
  \pmatrix{ \tilde{w}_0 \cr \vdots \cr \tilde{w}_{M-1} }\;.
\eeql{metricmat}
The necessary and sufficient condition is that the metric matrix on the
LHS is nonsingular. Singularity of this matrix would mean that a
nontrivial linear superposition $w$ of the $\tilde{w}_m$ (i.e., a
nonzero vector in $W$) is perpendicular to all $v_n$, i.e., that $w\in
V^\perp$. This simple calculation not only proves our assertion but, in
any finite-dimensional space, also gives a constructive algorithm.

Returning to the system of outgoing waves under discussion, the
contour-integral calculation of the main text now in effect shows that
(\ref{WVperp}) is indeed satisfied for $V_j={\cal
L}[\{\ket{f_{j,n}}\}_{n=0}^{M_j-1}]$ and $W_j={\cal F}V_j$, solves
(\ref{metricmat}) for this case, and extends this bi-orthogonalisation
to the whole Hilbert space by showing that the latter equals
$\oplus_jV_j$, with (\ref{blk-ortho}) holding between different
blocks.

 

\begin{references}
\bibitem{noDdagger}Conjugate linearity of the duality map at once shows
that $\prod{\psi}{{\cal D}\chi}=\prod{\chi}{{\cal D}\psi}$, so that
there is no independent role for a ${\cal D}^{\dagger}$.
\bibitem{c-norm}Stricty speaking, this symmetry is violated if the
normalization $\prod{f^j}{f_j}$ is not chosen real, as is for instance
the case in~\cite{bior}. However, this will cause no problems here.
\bibitem{c-spec}Since our NHH $H$ will operate on function pairs
defined on a finite interval (cf.\ below (\ref{eq:hwe})), we do not
need to consider the possibility of a non-eigenvector continuous
spectrum.
\bibitem{fnjordan}
Strictly speaking, Jordan blocks refer to a specific
way of representing these blocks; see, e.g., \cite{I&P}.  However, we
shall use this term here simply to mean
that $H$ cannot be fully diagonalized.
\bibitem{fncomp}
The property that the dynamics can be represented entirely by the
discrete poles has been referred to as ``completeness"; see, e.g.
\cite{comp1,comp2,tong}.  However, this nomenclature will be avoided in
the present paper, since the focus of attention will be precisely on
those cases where the dynamics can be so represented, but the set of
eigenvectors is nevertheless incomplete.
\bibitem{fn2}
This condition is appropriate for waves
emitted at some finite time in the past, and differs
from usual discussions of resonances.
\bibitem{simple} For open systems, the order of the zero of $W$
coincides with the order of the pole in $\tilde{G}$ not only for
generic $x,y$, but in fact for all $x,y>0$. Namely, a node
$f(x_0,\omega_j)=0$ or $g(x_0,\omega_j)\propto f(x_0,\omega_j)=0$ would
imply a vanishing energy current at position $x_0$, which is
incompatible with the dissipative nature of the QNMs. This observation
is also relevant to the discussion in Section~\ref{sect:dual}.
\bibitem{fnnotation}
Since the arguments of the bilinear map are always two-component
vectors we avoid the pedantry of $(\ket{\psi},\ket{\chi})$. In some of
our earlier papers, the bilinear map
is called a generalized inner product or norm,
and sometimes the notation $\prod{\psi}{\chi}$
is adopted for it.
\bibitem{fnrepul}
Time-independent perturbation theory can be formulated in a manner
similar to the conservative case; see, e.g., \cite{tong}.  Generically
this leads to level repulsion in one direction in the complex frequency
plane, and attraction in the other direction.  Thus the trajectories of
two eigenvalues would typically be like a pair of hyperbolas which come
close to each other but do not touch; see, e.g., the examples relating
to absorptive systems in \cite{comp1}.
\bibitem{j0} Following our conventions in previous papers
\cite{tong,sq}, QNMs on the negative imaginary $\omega$-axis will be
termed zero-modes and labeled with $j=0$ if necessary.
\bibitem{fnevenodd}
Here the Klein--Gordon equation is considered
on the full line $-\infty < x < \infty$
rather than the half-line $0 \le x <\infty$.
Only trivial changes are required in the formalism;
see, e.g., \cite{comp1,comp2,tail}.
\bibitem{vary-cst} That the $\ket{f_{j,n}}$ are linearly independent of
course does not mean that they are unrelated. In fact, by
differentiating $D(\omega)f(x,\omega)=0$ with respect to $\omega$ and
solving the ensuing $D(\omega_j)f_{j,1}=-2\omega_j\rho f_j$ by
variation of the constant, one finds (for $M\ge2$)
$f_{j,1}(x)=2\omega_jf_j(x)\int_x\!dy\,f_j^{-2}(y)
\int_0^y\!dz\,\rho(z)f_j^2(z)$,
where the undetermined constant of integration corresponds to the
freedom pointed out below (\ref{Hmat}). For $M\ge3$, higher derivatives
can be calculated in a similar fashion. However, the expressions in the
main text are more transparent if the $f_{j,n}$ are left unevaluated.
\bibitem{normW} One has
$W_{j,M_j+n}=\sum_{m=0}^nf'_{j,n-m}(0)g_{j,M_j+m}(0)$, and for $n\le
M_j-1$ this does {\em not\/} depend on the choice of $g_{j,M_j+m}$.
Namely, $g(x,\omega)\mapsto[1+(\omega-\omega_j)^{M_j}{\cal
N}(\omega)]g(x,\omega)$ yields $g_{j,M_j+m}\mapsto
g_{j,M_j+m}+\sum_{\ell=0}^m{\cal N}_{j,m-\ell}g_{j,\ell}$, with
$g_{j,\ell}(0)=0$ since $\ell\le M_j-1$. Thus (\ref{Wexp}) only depends
on the normalization of $f$ and $g$ up to order $M_j$, as stated in the
main text. In this context the reader is also referred to (\ref{fgW})
where, at the expense of a slightly more involved calculation, the
derivative $W_{j,n}$ (with $n\le2M_j-1$) is expressed in a form which
{\em manifestly\/} involves only $f_{j,m},g_{j,m}$ for $m\le M_j-1$.
\bibitem{fnmap}
In \cite{tong}, the duality transformation is not explicitly
mentioned.  Rather, bilinear maps $(\psi , \chi)$ are used; these can
be cast into the BB language by (\ref{eq:drelate}).
\bibitem{degenerate} Thus, we shall avoid the terminology
``degeneracy'' in this context.
\bibitem{rho-sing} Strictly speaking, $\rho^{-1}(x)$ is undefined if
$\rho$ contains a $\delta$-function as in (\ref{eq:ex1}). However, this
can always be dealt with as a limiting case of regular $\rho$. To be
sure, in the formulas of Section~\ref{degpt-ex} no problems arise.
\bibitem{numint} In more systematic numerical work, one should
circumvent the multiple integrals in (\ref{thirdO}) and
(\ref{fourthO}), which as they stand are very time-consuming. For
instance,
$p=4\int_0^1\!dx\,f^{-2}(x)\bigl[\int_0^x\!dy\,f''(y)f(y)\bigr]^2$ is
the solution $p(1)$ of the system $(p,q)'=(4q^2/f^2,f''f)$ with initial
conditions $(p(0),q(0))=(0,0)$.
\bibitem{rho0}In this example, $\rho(x)\rightarrow0$ if $x\rightarrow0$
and $n>3$, but this can always be remedied by adding a small positive
$x^3$ term and adjusting $\alpha$ accordingly.
\end{references}
\end{document}